\documentstyle[12pt]{article}
\def\>{\rangle}
\def\<{\langle}
\def\n{\noindent}
\def\|{|\!|}
\def\H{{\cal H}}

\def\{{\lbrace}
\def\}{\rbrace}

\def\E{{\rm E}\hskip-.53em{\rm I}\;}

\def\ra{\rightarrow}

\def\tint{{\textstyle{\int}}}
\def\bea{\begin{eqnarray}}
\def\eea{\end{eqnarray}}

\def\>{\rangle}
\def\<{\langle}
\def\|{|\!|}
\def\H{{\cal H}}

\title{Coherent States for \\ Discrete Spectrum Dynamics}
\date{}
\author{John R.~ Klauder\\
Department of Physics \footnote{Also Department of Mathematics.}, 
University of Florida\\
P.O. Box 118440, Gainesville, FL 32611, USA\\E-mail: klauder@phys.ufl.edu}

\begin{document}
\maketitle

\abstract{Coherent states for general systems with discrete spectrum,
such as the bound states of the hydrogen atom, are discussed. The states 
in question satisfy: (1)
continuity of labeling, (2) resolution of unity, (3) temporal stability, 
and (4) an action identity. This set of reasonable physical requirements 
uniquely specify coherent states for the (bound state portion of the) 
hydrogen atom. }

\section{Introduction}
Coherent states for the harmonic oscillator have long been known and their
properties have frequently been taken as models for defining coherent 
states for other systems. Interest has long existed in coherent states 
for the hydrogen atom, at least for the bound state portion, and there 
have been a number of attempts in the past to define such states. We 
shall approach this problem by adopting four postulates. The first two 
are standard when dealing with coherent states, while the third and 
fourth are rather more physical in nature dealing with a specific 
Hamiltonian operator. 
  
The first two postulates are given by: \vskip.1cm
\hskip1cm (1) {\it Continuity of labeling} \vskip.02cm
\hskip1cm (2) {\it Resolution of unity}\vskip.1cm

Indeed, these postulates are the minimal requirements generally 
accepted at present to characterize coherent states.~\cite{kla} The 
remaining two postulates refer to a specific Hamiltonian $\H\ge0$, 
which, in units where $\hbar=1$, we assume is nondegenerate and 
fulfills $\H\,|n\>=E_n\,|n\>=\omega e_n\,|n\>$.
For convenience we confine our attention to Hamiltonians with an 
infinite number of bound states, $0\le n< \infty$. 
We set $\lim_{n\ra\infty}E_n=E^*$; cases where $E^*<\infty$ and 
where $E^*=\infty$ are both of interest.
Note that the $e_n$ are dimensionless numbers which satisfy the 
property that
   $ 0=e_0<e_1<e_2<\cdots\,$.

The appropriate coherent states are defined by the expression
  $$|J,\gamma\>\equiv M(J)^{-1}\sum_{n=0}^\infty( J^{n/2}\,
e^{-ie_n\gamma}/\sqrt{\rho_n}\,)\;|n\>\;,$$
for suitable $J\ge0$ and $-\infty<\gamma<\infty$. 
The coefficients $\rho_n$ are chosen as the moments of a distribution
$\rho(J)\ge0$ in the manner
  $$\rho_n = \int_0^{J^*} J^n\rho(J)\,dJ\;,\hskip1cm \rho_0=1\;.$$
Here $M$ denotes a normalization constant chosen so that $\<J,\gamma|
J,\gamma\>=1$, namely $M(J)^2=\sum_{n=0}^\infty J^n/\rho_n$,
and we denote by $J^*$ the radius of convergence for this series. Cases 
where $J^*<\infty$ and $J^*=\infty$ are both of interest.
  
For the third postulate we choose\vskip.1cm
\hskip1cm (3) {\it Temporal stability}\vskip.1cm
\n which means that the time evolution of a coherent state remains a 
coherent state for all time. In particular, for the case at hand,
 $$  e^{-i\H t}|J,\gamma\>=\sum_{n=0}^\infty (J^{n/2}\,e^{-ie_n\gamma}/
\sqrt{\rho_n}\,)\,e^{-ie_n\omega t}\,|n\>=|J,\gamma+\omega t\>\;.$$
We observe that the expressions $J(t)=J$ and $\gamma(t)=\gamma+\omega t$ 
are reminiscent of the time dependence of action-angle variables in 
classical mechanics. When expressed in canonical action-angle variables, 
the classical action functional becomes
  $$I=\tint[J(t){\dot\gamma}(t)-\omega J(t)]\,dt\;. $$
Recall \cite{kla2,kla} that the quantum action functional restricted just 
to coherent states, namely
  $$I=\tint[\,i\<J(t),\gamma(t)|(d/dt)|J(t),\gamma(t)\>-\<J(t),\gamma(t)|
\H|J(t),\gamma(t)\>\,]\,dt\;,$$
also expresses the classical action. Thus our fourth postulate is 
the\vskip.1cm
\hskip1cm (4) {\it Action identity}\vskip.1cm
\n which ensures that $J$ and $\gamma$ are action-angle variables, and 
requires that
  $$\<J,\gamma|\H|J,\gamma\>=\omega J=\omega (\Sigma_ne_nJ^n/\rho_n)\,
/\,(\Sigma_mJ^m/\rho_m)\;. $$
To satisfy the action identity requires that 
   $$\rho_n=e_ne_{n-1}e_{n-2}\cdots e_1\;. $$

For the harmonic oscillator, $e_n=n$, $\rho_n=n!$, $E^*=J^*=\infty$, 
$\rho(J)=e^{-J}$, $M(J)^2=e^J$, and therefore
  $$|J,\gamma\>=e^{-J/2}\sum_{n=0}^\infty(J^{n/2}e^{-in\gamma}/
\sqrt{n!}\,)\,|n\>\equiv |z\>\;, $$
which is just a standard canonical coherent state with $z\equiv 
J^{1/2}e^{-i\gamma}$. In this case it suffices to choose 
$-\pi<\gamma\le\pi$.

We conclude with a brief discussion of a one-dimensional hydrogen-atom 
analog with spectrum $E_n=E_0-\omega/(n+1)^2$, i.e., 
$e_n=1-1/(n+1)^2$. It follows in this case that 
  $$|J,\gamma\>=M(J)^{-1}\sum_{n=0}^\infty(\sqrt{(2n+2)/(n+2)}\,
J^{n/2}e^{-i\gamma[1-1/(n+1)^2]})\,|n\>\;. $$
Here $$M(J)^2=[J(1-J)]^{-1}+J^{-2}\ln(1-J)\;,\hskip1cm 0\le J< J^*=1\;. $$
For the system at hand these states satisfy the four cited postulates 
presented in this paper.

In addition, for this example, it is noteworthy that 
   $$\<J,\gamma|\H^2|J,\gamma\>=\omega^2J^2+\omega^2 v(J)\;, $$
where $0<v(J)<6(1-J)$. Thus as $J\ra1$, $v(J)\ra0$, indicative of a 
very narrow distribution.

An earlier letter \cite{kl3} discussed the (bound state portion of the) 
hydrogen-atom coherent states along the lines of this note without, 
however, introducing the action identity. (Some other analyses of 
hydrogen-atom coherent states are referenced in that letter as well).
\section*{Acknowledgments}
The author thanks J.P.~Gazeau and K.~Penson for discussions. An 
extension of the present work to include continuous and mixed spectra 
has been submitted.~\cite{gaz}



\begin{thebibliography}{99}
\bibitem{kla} J.R. Klauder and B.-S. Skagerstam, ``Coherent States'', 
(World Scientific, Singapore, 1985).
\bibitem{kla2} J.R. Klauder, {\it J. Math. Phys.} {\bf 4}, 1058 (1963).
\bibitem{kl3} J.R. Klauder, {\it J. Phys. A: Math. Gen.} {\bf 29}, L293 
(1996).
\bibitem{gaz} J.P. Gazeau and J.R. Klauder, ``Coherent States for 
Systems with Discrete and Continuous Spectrum'', submitted for publication.
\end{thebibliography}
\end{document}